# Debiased albedo distribution for Near Earth Objects

A. Morbidelli[1], M. Delbo[1], M. Granvik[2], W. F. Bottke[3], R. Jedicke[4], B. Bolin[5], P. Michel[1], D. Vokrouhlicky[6]

[1]Laboratoire Lagrange, Université Côte d'Azur, CNRS, Observatoire de la Côte d'Azur, Nice, France
[2]Department of Physics, 00014 University of Helsinki, Finland and Division of Space Technology, University of Technology, 98128 Kiruna, Sweden
[3]Southwest Research Institute and NASA's SSERVI-ISET Team, Boulder, Co.
[4]Institute for Astronomy, University of Hawai'i, Honolulu, HI
[5]Department of Astronomy, University of Washington, Seattle, WA
[6]Institute of Astronomy, Charles University, Prague 8, Czech Republic

**Abstract**
We extend the most recent orbital – absolute magnitude Near Earth Object (NEO) model (Granvik et al., 2018) to provide a statistical description of NEO geometric albedos. Our model is calibrated on NEOWISE albedo data for the NEO population and reproduces these data very well once a simple model for the NEOWISE observational biases is applied. The results are consistent with previous estimates. There are ~1,000 NEOs with diameter D>1km and the mean albedo to convert absolute magnitude into diameter is 0.147. We don't find any statistically significant evidence that the albedo distribution of NEOs depends on NEO size. Instead, we find evidence that the disruption of NEOs at small perihelion distances found in Granvik et al. (2016) occurs preferentially for dark NEOs. The interval between km-sized bodies striking the Earth should occur on average once every 750,000 years. Low and high albedo NEOs are well mixed in orbital space, but a trend remains with higher albedo objects being at smaller semimajor axes and lower albedo objects more likely found at larger semimajor axes.

## 1. Introduction

Granvik et al. (2016, 2018) have recently proposed a new model of the orbital and absolute magnitude distribution of Near Earth Objects (NEO) that supersedes the previous model by Bottke et al. (2000, 2002). Although the two models were built on similar principles, the new one is more accurate for several reasons: (i) it was constructed from a much more extensive set of numerical simulations that tracked how main belt asteroids escape into NEO space, (ii) all numerical integrations of asteroid orbital evolution were performed with much shorter timesteps, (iii) the new NEO model was calibrated on ~7,000 discoveries or accidental re-discoveries of NEOs by the Catalina Sky Survey, rather than the 138 detections of the Spacewatch Survey used to calibrate the Bottke et al. (2002) model, (iv) it accounted for source-specific absolute-magnitude distributions.

The Bottke et al. model had been extended by Morbidelli et al. (2002) to achieve a statistical description of the albedo distribution of NEOs. Knowledge of the albedo distribution is required to convert the absolute magnitude distribution of NEOs, which is estimated by using the photometric data obtained by asteroid surveys, into a size frequency distribution. When this information was convolved with estimates of terrestrial planet impact probabilities as a function of NEO orbital parameters, Morbidelli et al. (2002) were able to assess the frequency of impacts on Earth larger than a given size; they used these data to objectively evaluate the impact hazard.

The present paper stands to the Granvik et al. (2018) model as the Morbidelli et al. (2002) paper stood to the Bottke et al. (2002) model. The goal is to update our NEO albedo distribution model in a way that is consistent with our new NEO orbital and absolute magnitude model. However, this is not just a matter of redoing the calculations of Morbidelli et al. (2002) with improved data. We have developed a new method to estimate the albedo distribution of NEOs that is in principle more accurate than the original one, as explained in section 2. In section 3 we will present some validity tests. The model is then applied in section 4 to convert the absolute magnitude distribution of Granvik et al. (2018) into a size distribution and to represent the mixing of NEOs of different albedos as a function of semi major axis (a), eccentricity (e) and inclination (i). We finally update the estimate of the frequency of impacts of NEOs larger than 1 km in diameter on the Earth. The conclusions will close the paper in section 5.

## 2. Building the NEO albedo model

The method of Morbidelli et al. (2002) was as follows. First, they divided the asteroids in 4 albedo classes as defined by Tedesco et al. (2002). Second, they estimated the fraction of asteroids in each class that were located within the approximate vicinity of the considered NEO source regions used by Bottke et al. (2002). Third, given that the contribution of each source to the overall NEO population was known from the Bottke et al. model, they propagated this information forward to assess the fraction of NEOs in each albedo class as a function of (*a,e,i*).

Although this procedure is correct in principle, it is in practice difficult to implement. Albedos for main belt asteroids are only known mostly for bodies considerably larger than typical NEO sizes (e.g., IRAS data mainly computed albedos for main belt asteroids that were larger than many tens of km in diameter; the WISE infrared survey is only complete for identified main belt asteroids that are approximately 7 km in diameter; Tedesco et al. 2002; Masiero et al. 2011).

Accordingly, Morbidelli et al. (2002) was forced to extrapolate the observed albedo distributions in the source regions to much smaller sizes. This step is not straightforward. Background asteroids and asteroid families often have different albedo and size frequency distributions, which in turn means the number of bodies within each defined albedo class in a considered region may change with size. Moreover, the size frequency distribution of both background and families are only modestly known for diameters near a few km, the observational limit of main belt asteroids, and are almost completely unknown at still smaller sizes. There is also the issue that small family members may have be located far enough from the family center and thus can be easily confused with the background. All of these considerations lead to a somewhat delicate extrapolation procedure that depended on a number of rather arbitrary assumptions.

Yet another concern was how Morbidelli et al. (2002) chose to define their NEO source regions in the main belt. The Bottke et al. and Granvik et al. models use the concept of "intermediate sources", defined as the most powerful resonances that cross the asteroid belt (the $\nu_6$ secular resonance, the 3:1 mean motion resonance with Jupiter etc.). These resonances, however, are mostly empty at any given time because the eccentricity of resonant asteroids evolves relatively quickly from small, main-belt like values, to large values typical of the NEO population (Gladman et al., 1997). From there, the bodies can encounter or impact various planets and satellites, they can be destroyed by a close approaches to the Sun (Granvik et al. 2016) or they can be ejected out of the solar system, usually through a Jupiter encounter. This rapid evolution means it is currently impossible to obtain the albedo distribution of resonant objects. The alternative is to characterize the albedos distribution of asteroids residing in main belt regions that continuously feed those resonances with new bodies. Most asteroids smaller than a few tens of km are believed to reach main belt resonances by the so-called Yarkovsky effect (Bottke et al., 2006), a thermal drift force that causes asteroids to migrate in semi major axes at different rates as a function of their orbits, sizes, spin vectors, and physical properties. The issue here is that the characteristic distance that a body of a given size has to travel to reach a chosen resonance in the main belt is unknown. If nothing destroys an asteroid, and it can maintain a steady course inward or outward, it is possible it could conceivably migrate across an entire section of the main belt (e.g., Granvik et al. 2017). This possibility makes it challenging to choose well-defined boundaries for each source region that is providing asteroids to a given resonance.

For all these reasons, we decided to abandon the approach of Morbidelli et al. (2002) and develop a new approach, now feasible thanks to the many hundreds of albedo measurements that have been made among NEOs from the NEOWISE survey (e.g., Mainzer et al. 2012).

In our revised method, we first divide the asteroids in three albedo ($p_v$) categories:

Category 1: $p_v \leq 0.1$

Category 2: $0.1 < p_v \leq 0.3$

Category 3: $0.3 < p_v$

where $p_v$ is the geometric albedo. We think that these three categories are more meaningful than those in Tedesco et al. (2002). In fact, Mainzer et al. (2012) (see for instance Fig. 3a in that paper) showed that the albedo distributions of PHAs and NEOs are quite flat inside the first and second category, whereas there is a discontinuity in albedo distribution at the 0.1 and 0.3 boundaries. Wright et al. (2016) also found a strong discontinuity at the 0.1 boundary and an exponential fall off above 0.3. Moreover, $p_v = 0.1$ is a practical boundary to distinguish with some confidence primitive asteroids of C, P and D spectroscopic classes from the other classes of objects (S, E, M etc.), although shock darkening can potentially smear this boundary between the types (Binzel et al., 2019).

The second more fundamental difference with the Morbidelli et al. (2002) approach is that we do not consider as an input an albedo distribution for each of the sources of the NEO population for the reasons explained above. Instead, we determine the albedo distribution in the source regions by fitting the resulting albedo distribution in NEO space to the albedo distribution observed by the NEOWISE survey, corrected for observational biases.

More precisely, the Granvik et al. (2016, 2018) NEO model is built from source regions that comprise three dynamically distinct populations (the Hungaria population, the Phocaea population, the dormant Jupiter family comets) and four main belt escape routes (the $\nu_6$ complex -inclusive of 4/1 and 7/2 mean motion resonances- the 3/1 mean motion resonance, the 5/2 mean motion resonance complex -inclusive of 8/3 and 7/3 resonances- and the 2/1 mean motion resonance complex -inclusive of 7/2 and 9/4 mean motion resonances and the z2 secular resonance). For each of the first 6 sources $s(1 \dots 6)$, we define a probability $p_s(c)$ that represents the fraction of the bodies coming from that source which belong to albedo category $c$ ($c = 1,2,3$). We assume that this probability is independent of the asteroid absolute magnitude $H$. We checked the validity of this assumption for the $\nu_6$ resonance, by considering each of its thirteen neighboring families – each with its own average albedo – and extrapolating their correlated $a$ and $H$ distributions to estimate the number of asteroids that each should have delivered to the resonance, as a function of absolute magnitude: for $16<H<24$ the fraction of delivered asteroids in albedo category 1 is constant within few percent. For each source, the probabilities $p_s(c)$ are gauged by the relationship $\sum_{c=(1,2,3)} p_s(c) = 1$. Moreover, we assume that all dormant Jupiter family comets are in albedo category 1, i.e. $p_7(1) = 1$, $p_7(2) = 0$, $p_7(3) = 0$. We stress that, because in the Granvik et al. model the contribution of each source to the overall NEO population changes with H, the albedo distribution of the resulting NEO population is in principle H-dependent even if those of the sources are not.

We consider a set of NEOs whose albedos have been measured by the NEOWISE survey. To illustrate the method, we first assume for simplicity that this dataset is not biased in albedo. Absence of bias in albedo means that, for an ideal set of objects all having the same orbits, absolute magnitudes and sky positions, the albedos measured by the NEOWISE survey would be representative of the true albedo distribution. This is not accurate, of course, but the correction for observational biases will be considered later.

For each NEO, identified by the index $n$, the orbital-magnitude NEO model provides the probabilities $P_s(n)$ to come from the sources s=1,..,7. The NEOWISE observations provide the albedo category $c_n$ of the object. According to the model, the probability that the object has an albedo category $c=c_n$ (equivalently, the probability that the model reproduces the albedo observation for asteroid $n$) is:

$\mathcal{P}(c_n)=\sum_s P_s(n)p_s(c_n)$  (1)

Considering all NEOs together, the probability that the model reproduces the albedos of NEOs determined by the NEOWISE survey (still assuming no biases – see below) is:

$L=\Pi_n \mathcal{P}(c_n)$

where $\Pi_n$ denotes the product over $n$. Thus, the parameters $p_s(c)$ (a set of 12 parameters, 2 for each of the 6 asteroidal sources) can be determined by maximizing the likelihood function (maximum likelihood method):

$Ł=\sum_n \mathrm{Log}[\mathcal{P}(c_n)]$  (2)

Like all surveys, the NEOWISE dataset has observational selection effects that must be properly accounted for. For a body with a given absolute magnitude $H$, the more its albedo $p_v$ increases, the fainter its infrared apparent magnitude $W3$ becomes (essentially because the object is smaller, but also because of the effect illustrated in Fig. 1). According to Mainzer et al. (2011), the NEOWISE detection efficiency function is close to 100% up to $W3$=9.5, drops to 50% at W3=10 and it is close to zero beyond $W3$=10.5. This means that asteroids with large enough albedos would not have been observed by the NEOWISE survey. Consequently, the NEOWISE database is skewed towards low albedos (both for a $H$ limited sample and – although at a lesser extent – for a size-limited sample).

We correct for this bias in the following manner. We assume that the limiting magnitude for NEOWISE is $W3_{lim} = 10$. For each NEO with a NEOWISE-determined albedo $p_v$, we look for all the reported observations at different epochs. We select the lowest apparent magnitude $W3_{min}$ among all reported values. If the object has $W3_{min}$ fainter than $W3_{lim}$ we reject the object; otherwise we consider it. This leaves us with 328 NEOs. For these, we compute the maximal albedo $p_v^{lim}$ that would have allowed detection by NEOWISE as follows.

We fix the asteroid's absolute magnitude $H$ and physical properties (the beaming parameter η, spin period and pole orientation etc.) to be equal to the observed values. In this way, the infrared flux $F_{IR}$ depends only on the albedo $p_v$, and is proportional to:

$F_{IR} \sim (1 - \alpha p_v)D^2$  (3)

where D is the diameter of the asteroid. The term (1 - $\alpha p_v$) is due to the fact that the thermal IR flux depends also on the asteroid temperature and not only on the body's surface area ($\propto D^2$). The surface temperature is proportional to $(1-qp_v)^{0.25}$ where q is the so-called phase integral. In order to determine the dependence of the thermal infrared flux on $p_v$, one needs in principle to calculate ∂$F_{IR}$/∂$p_v$. This derivative can be quite complicated, such that we used numerical methods to compute the effect of varying $p_v$ on $F_{IR}$. Using the NEATM model (Harris, 1998), we calculated the $F_{IR}$ for a 1 km diameter spherical asteroid located at 1, 1.5, and 2 AU from the Sun. The phase angle was set to 30° and we changed $p_v$ from 0 to 1. We normalized these fluxes to the value of $F_{IR}$ for $p_v=0$. The results are shown in Fig. 1, which shows that $F_{IR} \propto (1 - \alpha\, p_v)$, where the coefficient α depends on the distance of the asteroid from the Sun.

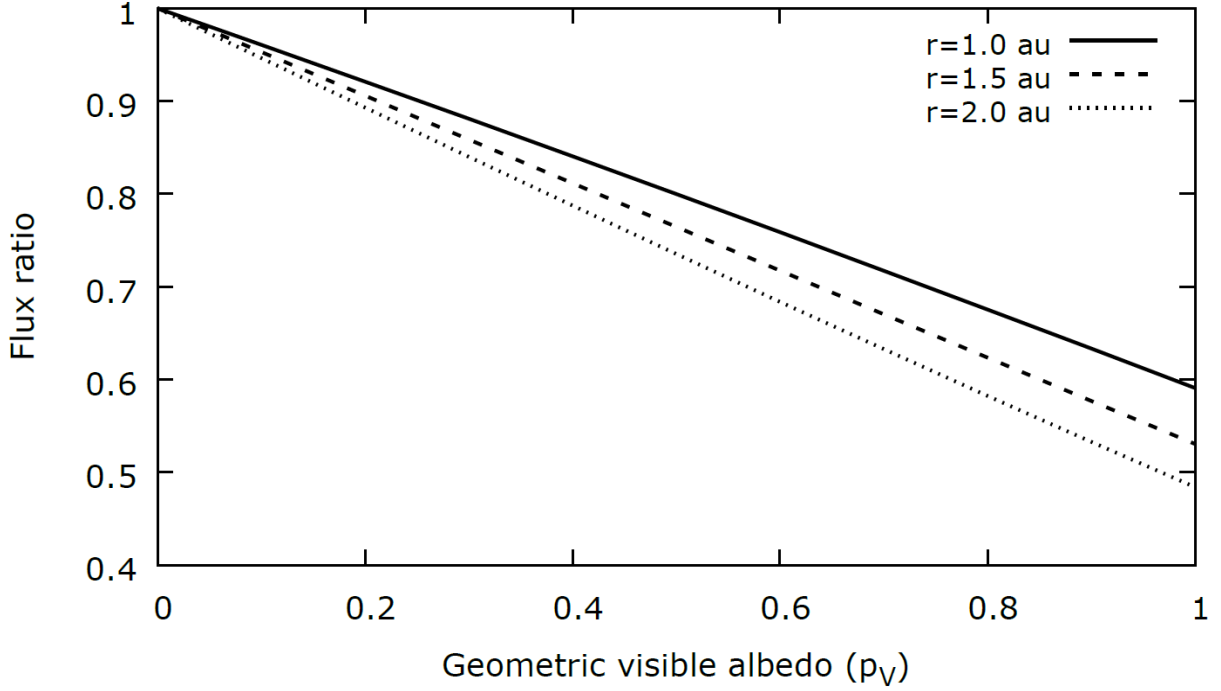

Fig. 1. The dependence of the IR flux $F_{IR}$ on the geometric visible albedo $p_v$, normalized by the value of $F_{IR}$ for $p_v=0$ for asteroids at a distance from the Sun of 1 AU (solid line), 1.5 AU (dashed) and 2 AU (dotted).

From the results of Fig. 1, we adopt the assumption $\alpha = 0.45$. The square of the diameter $D$ is proportional to the light in the visible band and inversely proportional to $p_v$.

$$D^2 \sim 10^{-0.4\,H}/p_v$$

Injecting this dependence in formula (3) and imposing $F_{IR}^{lim} = 2.512^{(W3_{min} - W3_{lim})} F_{IR}$, where $F_{IR}$ is the infrared flux corresponding to magnitude $W3_{lim}$, we find:

$$p_v^{lim} = \left[2.512^{(W3_{min} - W3_{lim})} \cdot (1. - 0.45\, p_v)/p_v + 0.45\right]^{-1}$$

To summarize, $p_v^{lim}$ is the value of the albedo that would have given the body (of absolute magnitude $H$) a magnitude $W3_{lim}$ at the time of its observation (when instead it had a magnitude $W3_{min}$ thanks to its real albedo $p_v$). Given the value $p_v^{lim}$ for each asteroid $n$, we can determine a bias $B_n(c)$, defined as the probability that that asteroid would have been observed if it had been in albedo category $c$. For this purpose we assume, from Mainzer et al. (2012), that the albedo distributions inside category 1 and 2 are uniform, while the albedo distribution in category 3 decays as:

$$N(p_v)\, dp_v = \left(\frac{1}{2.6}\right)^{(p_v - 0.3)/0.1} dp_v \qquad (4)$$

(i.e. a factor of 2.6 in each 0.1 wide albedo bin for $p_v > 0.3$) and define:

$$B_n(1) = \min\left(\frac{p_v^{lim}}{0.1}, 1\right)$$

$$B_n(2) = \min\left(\frac{p_v^{lim} - 0.1}{0.2}, 1\right)$$

$$B_n(3) = \min\left(\frac{1 - e^{-log(2.6)\frac{p_v^{lim} - 0.3}{0.1}}}{1 - e^{-log(2.6)\frac{1 - 0.3}{0.1}}}, 1\right)$$

The more complex albedo distribution from Wright et al. (2016) could also be adopted but we find it is sufficiently well approximated by our assumption of a flat (but different) distribution in the 0-0.1 and 0.1-0.3 intervals and an exponential fall-off beyond 0.3.

With this definition of the biases, formula (1) above now becomes:

$$\mathcal{P}(c_n) = \sum_s \frac{P_s(n)\big(B_n(c_n)p_s(c_n)\big)}{B_n(1)p_s(1) + B_n(2)p_s(2) + B_n(3)p_s(3)} \quad (5)$$

This new definition of $\mathcal{P}(c_n)$ is then applied in formula (2) for the computation of the parameters $p_s(c)$ by the maximum likelihood method. The Likelihood function (2) is maximized using the amoeba routine (Press et al., 1992) starting from a system of 1,000 simplexes, so that the values of the best fit coefficients $p_s(c)$ are retrieved.

In order to estimate the uncertainty of the determined coefficients $p(c)$, we repeated this procedure assuming $W3_{lim}$=9.5, which reduces the number of NEOs to 232. We have taken as nominal values for the coefficients $p_s(c)$ the average of the values obtained assuming $W3_{lim}$=9.5 and $W3_{lim}$=10. Our error bar is defined as the half difference between the two.

Table 1 reports the results. It shows, as expected, that the fraction of asteroids in category 1 (low albedos) is relatively small for inner belt sources (Hungaria, $\nu_6$, 3/1) and relatively large for outer belt sources (5/2 and 2/1). It is interesting to note that a large fraction of low-albedo asteroids are predicted to exist in the Phocaea region. We shall examine this value in more detail below.

Now that the albedo distribution in each source has been obtained, the albedo distribution in each (a,e,i) cell of the NEO model is obtained by combining the albedo distribution of each source with the relative contribution of the sources to the NEO population in the considered cell. The relative contribution of the sources, as well as the uncertainties, are provided by the NEO model of Granvik et al. (2018).

Table 1: albedo probabilities for the 6 asteroidal sources of NEOs. Notice that $p_s(3)$=1-($p_s(2)$+ $p_s(2)$).

| source | $p_s(1)$ | $p_s(2)$ | $p_s(3)$ |
| --- | --- | --- | --- |
| $\nu_6$ | 0.120 +/- 0.014 | 0.558 +/- 0.003 | 0.322 +/- 0.014 |
| 3/1 | 0.144 +/- 0.034 | 0.782 +/- 0.036 | 0.074 +/- 0.050 |
| Hungaria | 0.021 +/- 0.005 | 0.113 +/- 0.004 | 0.866 +/- 0.006 |
| Phocea | 0.501 +/- 0.010 | 0.452 +/- 0.020 | 0.047 +/- 0.022 |
| 5/2 | 0.294 +/- 0.047 | 0.557 +/- 0.039 | 0.149 +/- 0.061 |
| 2/1 | 0.399 +/- 0.015 | 0.200 +/- 0.036 | 0.461 +/- 0.039 |

## 3. Validity Tests

As a direct test of the accuracy of our NEO albedo model, we compare its predictions with the data that we attempted to fit. To do this, we divided all NEOs with NEOWISE-derived albedos into binned one-dimensional distributions in semi major axis, perihelion distance and inclination. This projection of the population into one-dimensional distributions is needed to make a meaningful comparison because we are only considering albedo data for 328 NEOs; it would be too sparse over a three-dimensional distribution. In each bin of each distribution, we computed the fraction of the NEOs that are in each albedo category.

This method yields the black histograms reported in Figures 2, 3 and 4. The error bar is computed assuming Poissonian statistics; namely, if N objects are in a given albedo category in an orbital element bin, we assume that the uncertainty is N1/2. The resulting error bars are shown as black vertical segments near the center of each bin.

For the model, we sum the values of $\mathcal{P}(c_n)$ given by (5) for all the asteroids in each orbital element bin (which are not necessarily the same because, for a given value of one orbital parameter -say the semi major axis- asteroids can have different eccentricities and inclinations and therefore have different source probabilities), with $c_n=1$ and $c_n=3$. This tells us which fraction of the population that is expected to exist and to be detected in each albedo category. The lower red histograms in Figs. 2-4 show the result for $c_n=1$ (i.e the fraction of NEOs with has $p_v$<0.1). The upper red histograms show the complement of the fraction of the population with $c_n=3$, i.e. the fraction of NEOs with has $p_v$<0.3). The model error bars are computed by propagating the uncertainties on the coefficients $p_s(1)$, $p_s(2)$ and $p_s(3)$ for all sources s, and are shown by the red vertical bars plotted near the center of each bin.

The agreement between model and observations is very good. One has to remember that the model fit is done on the individual NEOs with albedos determined by NEOWISE and not on the observed binned 1-dimensional distributions. Thus, the model is strongly under-parametrized (12 parameters) compared to the number of data to be fit (hundreds of NEOs), so that obtaining visual matches with the observed distributions as good as those shown in Figs. 2-4 is non-trivial.

In the semi major axis distribution, both data and model show the decrease of the fraction of low-albedo asteroids (category 1) from ~90 to ~15% with decreasing semi major axis, whereas the fraction of objects with $p_v$<0.3 decreases only from 90 to 80% (Fig. 2). For inclinations, the albedo distribution seems to be relatively flat (Fig. 3).

We find the most interesting feature to be the albedo distribution for perihelion distance $q$ (Fig. 4). The fraction of low-albedo asteroids only declines moderately (from ~50% to ~30%) as $q$ goes from 1.3 AU to 0.3 AU. In the $q < 0.2$ AU bin, however, there are no low-albedo NEOs in NEOWISE database, whereas the model remains at the 30% level.

The absence of low-albedo low-$q$ NEOs was originally pointed out in Mainzer et al. (2012), who interpreted this as evidence that low-$q$ NEOs come from a source regions exclusively inhabited by high-albedo asteroids. The fact that the model and the data do not agree in that bin (and exclusively in that bin) suggests instead a different interpretation, namely the physical disruption of dark asteroids at low perihelion. This scenario was argued by Granvik et al. (2016), who found a deficit of $q < 0.7$ AU NEOs relative to expectations based from their orbital-magnitude model. They interpreted this observation as evidence that many asteroids disintegrate when they acquire a perihelion distance smaller than ~ 0.1 AU. The disappearance of low-albedo NEOs in the smallest $q$-bin implies that the physical disruption mechanism predominantly affects carbonaceous chondrite-like asteroids. While this mechanism has yet to be fully characterized, it seems likely that such bodies are more easily destroyed close to the Sun because of their greater abundances of volatiles and hydrated material and/or their higher vulnerability to thermal cracking (Delbo et al., 2014).

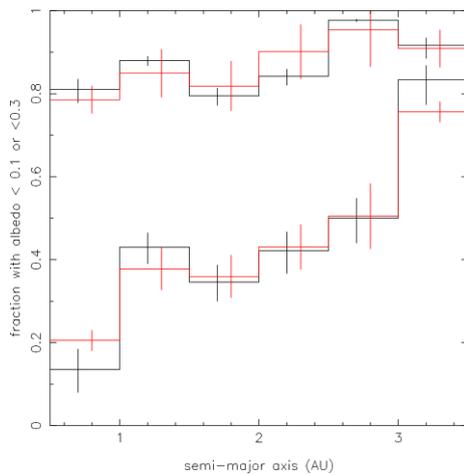

Fig.2: The fraction of the NEO population that has $p_v$<0.1 (lower histograms – category 1) and $p_v$<0.3 (upper histograms – category 1+2), binned in semi major axis. Category 3 is represented by the complement to unity of the upper histograms. The black histograms are computed from the albedos of NEOs observed by NEOWISE. The red histograms are computed from our albedo distribution model and taking into account the NEOWISE albedo bias for each observed object, as explained in the text. Notice that model and data agree in all bins within the error bars.

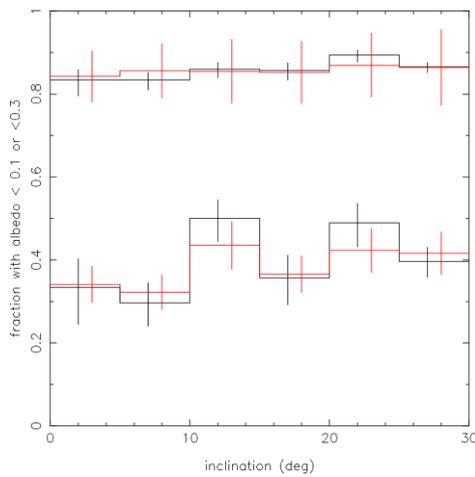

Fig. 3: The same as Fig. 2 but for the NEOs binned in inclination.

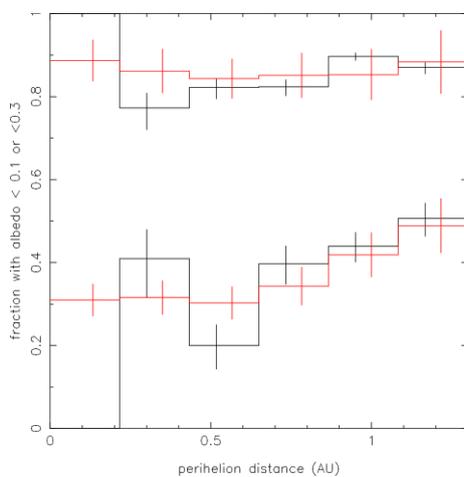

Fig. 4: The same as Fig. 2 but for the NEOs binned in perihelion distance.

Like Granvik et al. (2018) introduced an empirical correction to the NEO orbital-magnitude model to account for the deficit of low-$q$ NEOs, we will also introduce an empirical correction to our albedo model, setting all NEOs with $q$<0.2 AU into category 2 in the model distribution that is made available to the community.

As a second test of the validity of the model, we can look at the albedo distributions predicted by the model inside the NEO source regions (Table 1). In particular, we checked the albedo distributions inside the Hungaria and Phocaea source regions. We choose these two because they are nicely bound in orbital element space, whereas it is more difficult to define which asteroids are the most predominant contributors to other sources like $\nu_6$ and 3/1. Moreover Hungarias and Phoceas have diagnostic features, according to Table 1. In the Hungarias, as much as 86 ± 8% of the asteroids should be in albedo category 3, unique among all sources. In the Phoceas, 50% of the asteroids should be in category 1.

To determine the "real" albedo distributions in these sources, we selected all asteroids in these two groups with WISE-determined albedos. The same procedure used for the NEOs was applied, but in this case we assumed that all probabilities $P_s$ are null except the one corresponding to the source under examination. We assume W3$_{lim}$=10.

For the Hungarias, we find $p_3(1)$= 0.052 and $p_3(2)$= 0.157 which are in quite good agreement with the values determined via the NEO model. In particular, the prediction that the overwhelming majority of asteroids are in category 3 (86 ± 8% according to the model) is confirmed: we find 80% among the Hungaria asteroids.

For the Phoceas, the result is less good, at least at first glance. Even if 70% of the Phocaea asteroids observed by WISE with W3<W3$_{lim}$ are in category 1, when we take into account the biases we find $p_4(1)$=0.23 and $p_4(2)$=0.36. These values are significantly lower than those reported in Table 1 from our model.

It is possible this difference in the Phocaea population is not diagnostic of a problem in our NEO albedo model. Consider that Novakovic et al. (2017) found a small dark asteroid family in the Phocaea region that, by its orbital position near an escape route, could potentially dominate the Phoceas' flux to the NEO population. This situation may make it possible for Phocaea-derived NEOs to not be representative of the overall Phocaea population.

It is also possible, however, that the predictive power of our model for the Phocaea region is simply limited. Only a small fraction of NEOs are expected to come from the Phoceas, so our model constraints for this region are modest at best. As evidence, consider that of the 328 NEOs with NEOWISE-derived albedos in our test set ($H$>15 and $W$3<10), only 18 are expected to come from the Phoceas. Thus, even if the albedo distributions in this source were not determined with great accuracy, the overall NEO-albedo distribution model would not be substantially affected.

## 4. Results

We now use our model to convert Granvik et al. (2018) $H$-distribution of NEOs into a size distribution and to map the albedo distribution in ($a,e,i$) orbital space.

### 4.1 Size distribution

In order to compute the distribution of NEOs as a function of size and its uncertainty, we operate as follows. From the model of Granvik et al. (2018), which is a probability distribution, we generate a population of synthetic NEOs in ($a,e,i,H$) space. For each synthetic NEO, which belongs to a cell in ($a,e,i,H$), the present work provides the probabilities that the object falls in each of the three albedo categories, and their uncertainties. From a random sampling of Gaussian distributions with widths equal to these uncertainties, we generate the actual probabilities. From these probabilities, another random choice attributes to the synthetic NEO an albedo category. A final random choice attributes to the NEO a value $p_v$ of the albedo according to the albedo distribution inside the selected category, described in Section 2 (a flat distribution for categories 1 and 2 and eq. (4) for category 3). Given $p_v$ the absolute magnitude *H* of the synthetic NEO is converted into a size *D* with the usual formula

$$D = 10^{[6.244-0.4H-\log(p_v)]/2} \tag{6}$$

From all the generated diameters of the synthetic NEOs we generate a cumulative size distribution. We repeat this procedure 100 times, each time getting a different NEO synthetic population due to the random choices described above, and from the 100 size distributions we get a mean distribution and its 1-σ uncertainty. Computing a variance of a cumulative distribution is notoriously complicated, because the datapoints of a cumulative distribution are correlated. To deal with this, we proceed as follows. For each synthetic NEOs population, we rank the objects by decreasing size, and we enumerate them from 1 to *N*, where *N* is the total number of generated NEOs (which changes for each synthetic population, from the uncertainty on the total number of NEOs with H<25 in Granvik et al. , 2018). Then we take a number *M*<$N_{min}$, where $N_{min}$ is the minimal value of *N* over the 100 generated NEO populations. From these populations we get the diameters $D_1$,….$D_{100}$ of the asteroids ranked number *M* in their respective distributions. The mean of these values will be the value of the diameter of the mean cumulative distribution for the *M*-th object, and the variance of the $D_1$,….$D_{100}$ values will give the 1-σ uncertainty on the mean value.

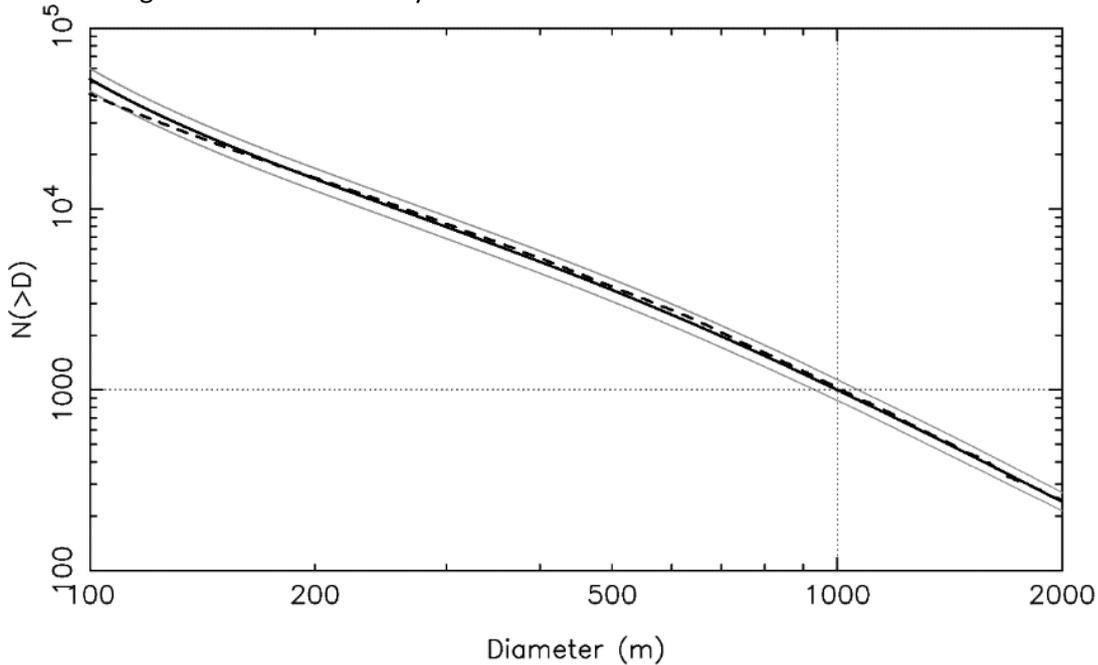

Fig. 5. The solid black curve shows the NEO size distribution obtained from Granvik et al. (2018) model, using the albedo distribution in each (*a,e,i,H*) cell. The two thin gray curves show the 1-σ uncertainty on the size distribution. The dash-black curve shows the size distribution obtained from the NEO *H*-distribution of Granvik et al. (2018), adopting a mean albedo of 0.147. This value has been chosen so that the solid black and gray curves match at *D*=1km. The two dotted lines mark the size D=1km and N(>1km)=1,000 for reference.

The resulting size distribution is presented in Fig. 5. It is interesting to note that the curves do not match at all sizes. They do not have to do so because the Granvik et al. model gives us different H-distributions for the NEOs coming from different sources and each source has a different albedo distribution. As a result, the albedo distribution of NEOs is H-dependent. The difference between the solid and dashed curves, however, is never very big: the dashed curve falls within the 1-σ uncertainty of the solid curve. Thus, the H-dependency of the mean albedo of NEOs is not statistically significant.

For reference, Pravec et al. (2012) found a mean albedo of 0.197 for main belt asteroids of S/A/L type and 0.057 for C/G/B/F/P/D types. This implies that most of NEOs for a given absolute magnitude are S-type, as expected because for a given H they are smaller and therefore more numerous.

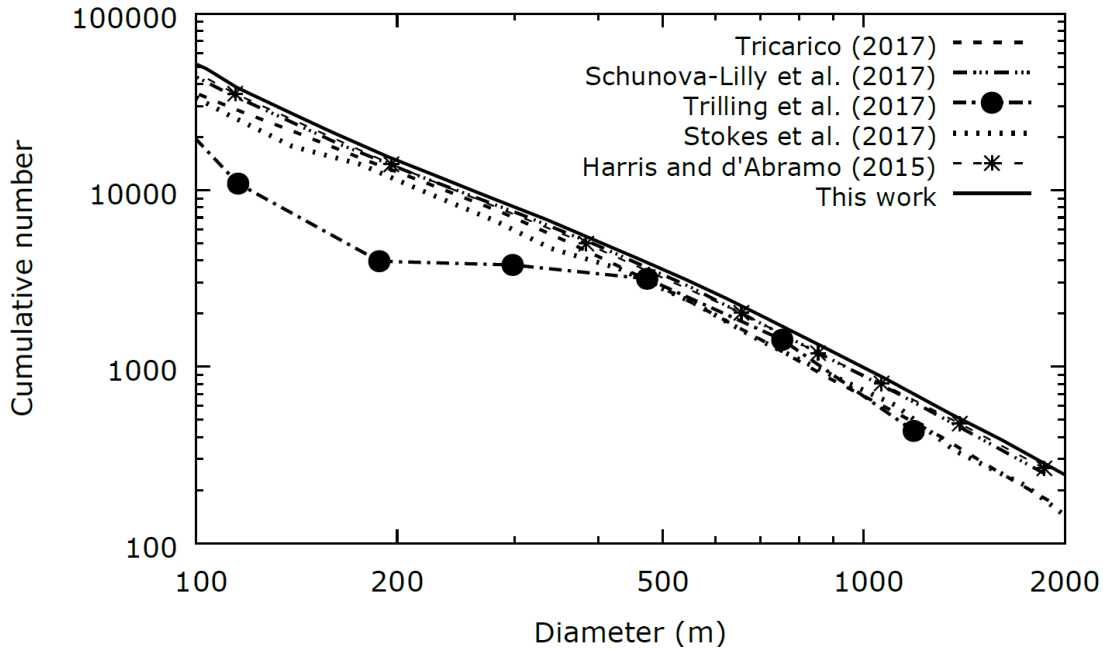

Fig. 6. Comparison among the size distributions of NEOs obtained in different works.

Fig. 6 compares the size frequency distribution that we obtain with those obtained by other authors. Our distribution predicts slightly more NEOs but essentially agrees with those of Harris and D'Abramo (2015), Stokes et al. (2017), Schunova-Lilly et al. (2017) and Tricarico (2017). In contrast, the size frequency distribution inferred by Trilling et al (2017), obtained from a limited number of objects (235 NEOs), shows an increasing mismatch starting at D < 500 m that becomes largest near 200 m. For 100 m bodies, we find a deficit of a factor of ~5 between their results and ours.

### 4.2 Albedo distribution and mixing in (*a,e,i*) space

The albedos of the largest 20,000 model NEOs as a function of their orbital parameters *a,e,i* is shown in Fig. 7 (i.e. down to a size of D = 190 m). The most striking information in this figure is the vast amount of mixing among NEOs of different albedo categories. Nevertheless, there is a general trend. NEOs of albedo categories 2 and 3 are more abundant at small semi major axis (*a*<2 AU), whereas low albedo NEOs of category 1 are predominant for *a*>2.5 AU. Both the mixture and the general semi major axis trend are indeed visible in the data and are quantitatively reproduced by our model (see Fig. 2 which, unlike Fig. 7, accounts for observational biases).

Note that in Sec. 4.1, we reported that the albedo distribution of NEOs is size-independent within statistical errors. This implies that adding additional smaller objects to Fig. 6 would not change the nature or trends in the figure in a substantial manner.

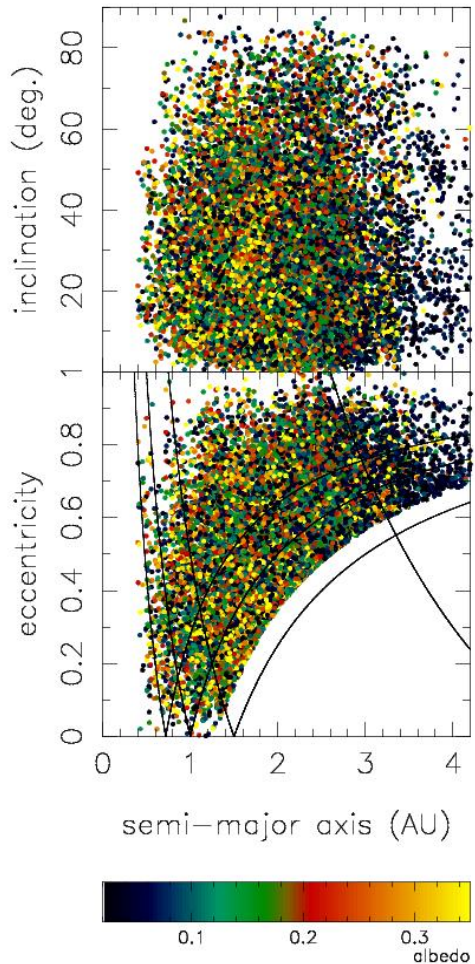

Fig. 7. The albedos of the largest 20,000 NEOs according to our model, as a function of the orbital parameters *a,e,i*. The solid curves mark the boundaries of the regions where NEOs cross the orbit of Venus, the Earth, Mars and Jupiter, each set on a zero eccentricity orbit with their current semi major axis, for simplicity.

### 4.3 Impact frequency of D>1km asteroids

In order to estimate the collision probability of NEOs with the Earth, we have used a new algorithm that was presented in Vokrouhlicky et al. (2012) and Pokorny and Vokrouhlicky (2013). It improves upon the traditionally-used Opik-like algorithm (Wetherill, 1967) in a number of ways. The most important change is there is no assumption that *a,e,i* are constant while the orbital angles undergo uniform precession. Instead, this algorithm accounts for how the evolution of *e* and *i* are coupled with the precession of the argument of perihelion $\omega$ due to the Lidov-Kozai cycle (Lidov, 1963; Kozai, 1962) induced by the encountering planet.

We assume the Earth's orbit is fixed, with $a_E$=1 AU and $e_E$=0.016. For the NEOs, we consider the values of *a,e,i* that characterize the centers of the cells of the Granvik et al. (2018) NEO model. We consider that *i* is measured relative to the ecliptic (so that $i_E$=0). For each set of (*a,e,i*), we consider 36 values of $\omega$ equally spaced in [0, 360) degrees and for each we have computed the collision probability with the Earth using the code provided to us by P. Pokorny (personal communication). The 36 results are averaged to get a mean collision probability characteristic of the considered cell. Finally, we average the collision probabilities obtained on all the cells of the NEO model, weighted by the fraction of the *H*<17.7 NEO population (corresponding to *D* > 1 km for a mean albedo of 0.147) that they contain.

We find a mean collision probability with the Earth per NEO of $1.33\times10^{-9}$ per year. This value folds in fact that Amor-type NEOs, defined as bodies not on Earth-crossing orbits, have zero probability of striking the Earth. Given that there are ~1,000 NEOs with $D$>1km (Fig. 5), we predict that the mean interval between collisions of km-size NEOs with the Earth is 754,000 years. This value is in line with previous predictions (e.g. Morbidelli et al., 2002).

We predict that $D$ > 100m bodies (~50,000 of them, according to Fig. 5) would collide with the Earth only once every 15,000 years. Clearly, the NEO impact hazard is very small.

## 5. Conclusions

We have extended the model of Granvik et al. (2018), describing the statistical distribution of NEOs in orbital space ($a,e,i$) and absolute magnitude $H$, to achieve a statistical description of the albedo distribution of NEOs. For this purpose, we have developed a new method that takes into account the albedos of NEOs measured by the NEOWISE survey, with an approximate estimate of the NEOWISE survey biases. The comparison between the model prediction and the NEOWISE observations is very good once these biases are taken into account. We continue to see compelling evidence for the disruption of NEOs at low perihelion distance, as argued in Granvik et al. (2016), and we show a preference for the destruction of low-albedo bodies.

Our results also show that low and high albedo NEOs are well mixed in NEO space, but a trend remains, with high albedo NEOs dominating at $a$ < 2 AU and low albedo NEOs dominating at $a$ > 2.5 AU.

Although in principle the albedo distribution of NEOs could be size-dependent, we do not find any statistical significant evidence for this dependency. Instead, we find that the total number of NEOs larger than a given size $D$ can be effectively computed from the total number of NEOs brighter than a given absolute magnitude $H$ by converting $H$ to $D$ using equation (6) and a mean albedo $p_v$=0.147. We predict the number of NEOs with $D \geq$ 1km is ~1,000 and the number of NEOs larger than 100 m is ~50,000 (see Fig. 5, also reporting the uncertainty on these population estimates). Our size frequency distribution is in very good agreement with that of Harris and D'Abramo (2015) and Stokes et al. (2017). We do not find support for substantial inflection points at 200-300 m, as suggested by Trilling et al. (2017).

Using our size frequency distribution and the computation of collision probabilities with the Earth, we find that km-size NEOs should collide with our planet once every 750,000 year on average. These results are consistent with previous work (Morbidelli et al. 2002). Based on NEO discovery statistics, we agree with Stokes et al. (2017) that the vast majority of km-sized impactors have been identified.

The fact that our results are consistent with previous work should be taken as an indication that our knowledge of the NEO population in terms of the orbits and sizes of $D$ > 100 m bodies is reasonably good. The full orbital- absolute magnitude – albedo – collision probability – impact velocity model of NEOs can be freely downloaded from http://neo.ssa.esa.int/neo-population.